\documentstyle[aps,prb,twocolumn,graphics]{revtex}

\begin{document}

\title{Conductance fluctuations and boundary conditions}
\author{Marc R{\"{u}}hl{\"{a}}nder$^1$, Peter Marko{\v{s}}$^{1,2}$
and C. M. Soukoulis$^{1,3}$ \\
$^1$Ames Laboratory and Department of Physics and Astronomy,
Iowa State University, Ames, Iowa 50011\\
$^2$Institute of Physics, Slovak Academy of Sciences,
D{\'{u}}bravsk{\'{a}} cesta 9, 842 28 Bratislava, Slovakia\\
$^3$Research Center of Crete, FORTH,
71110 Heraklion, Crete, Greece%\\[0.1in]
%\parbox[t]{5.5in}{\small
%The conductance fluctuations for various types for two-- and 
%three--dimensional disordered systems with hard wall and periodic
%boundary conditions are studied, all the way from the ballistic (metallic)
%regime  to the localized regime. It is shown that the universal conductance
%fluctuations (UCF) depend on the boundary conditions. The same holds
%for the metal to insulator transition. The conditions for observing 
%the UCF are also given.\\[0.1in]
%PACS numbers: 71.30.+h, 71.55.Jv}
}
\maketitle

\begin{abstract}
The conductance fluctuations for various types for two-- and 
three--dimensional disordered systems with hard wall and periodic
boundary conditions are studied, all the way from the ballistic (metallic)
regime  to the localized regime. It is shown that the universal conductance
fluctuations (UCF) depend on the boundary conditions. The same holds
for the metal to insulator transition. The conditions for observing 
the UCF are also given.
\end{abstract}


The influence of the boundary conditions (bc) on critical phenomena
in disordered mesoscopic systems has been demonstrated by studies
of the conductance distribution\cite{Sou99a,Sle00,Bra01}
and energy level statistics.\cite{Bra97}
The ensemble average of the logarithm of the conductance, $\left\langle
\ln(g)\right\rangle$, is smaller for hard wall boundary conditions than for
periodic boundary conditions. The variance $\left\langle\ln^2(g)
\right\rangle - \left\langle\ln(g)\right\rangle^2$ on the other
hand is larger for hard wall than for periodic boundary conditions.
The distribution of nearest neighbour energy level separations
$P(s)$ becomes more Wigner--Dyson--like for periodic boundary conditions.
Thus, systems with periodic boundary conditions exhibit a ``more metallic''
behaviour than those with hard wall boundary conditions.
Different boundary conditions also lead to different values
for the universal conductance fluctuations (UCF) in the diffusive metallic 
regime where the mean free path $l$
is much smaller and the localization
length $\xi$ much bigger than the system size.
The variation of ensemble fluctuations of the conductance
as the disorder increases throughout the metallic regime has been
studied,\cite{Hig92,Tan94}
but the influence of the boundary conditions in the ballistic regime where $l$
exceeds the system size, and close to the localized regime where
$\xi$ becomes comparable with the system size, has not been studied
in detail. Also, the role of the correlation length in 
samples with a true metal--insulator--transition (MIT) --- i.e. in
systems where the localization length is ``infinite'' in the
metallic regime --- has not been discussed. We present here
numerical studies of cubic systems and squares with spin--orbit
scattering --- all of which have a true MIT --- for both
hard wall and periodic boundary conditions in the direction(s)
perpendicular to transport.

We are using the tight--binding model with the Hamiltonian

\begin{equation}
{\mathcal H} = \sum_{n,\tau} \left| n\tau \right\rangle \varepsilon_n
\left\langle n\tau \right| + \sum_{n,\tau,n',\tau'} \left| n\tau
\right\rangle V_{n,n'} \left\langle n'\tau' \right|
\end{equation}

\noindent
where $n,n'$ are nearest neighbour lattice sites on a square or cubic
lattice. For systems with spin--orbit--interactions $\tau$ and $\tau'$
take on values of $+1$ or $-1$ and the hopping integrals $V_{n,n'}$ are
$2\times 2$ matrices; without spin--orbit--interactions, the hopping
integrals are scalar and the spin ``variables'' have only one value.
We take the site energies $\varepsilon_n$ (independent of $\tau$)
to be random variables, chosen from an interval $\left[ -W/2; W/2 \right]$
with a uniform probability distribution. The parameter $W$ serves 
thus as a measure of disorder strength. 
The conductance is calculated using the transfer matrix method and
the Landauer formula.\cite{Eco81}

We have used the analytical derivation of Lee et al.\cite{Lee85} to calculate
theoretical values for the UCF for both types
of boundary conditions. 
In order to change the boundary conditions to periodic, one needs
to make the following changes (references to equations are from the
Appendix of Lee et al. (1987)\cite{Lee85}):
in the eigenfunctions $Q_m$ to the diffusion equations, the
cosines in the transverse directions (Eq.\ (A9)) must be replaced by
exponentials with a factor of $2\pi$ instead of $\pi$ in the argument;
this will lead in effect to a factor $4$ in the $m_x^2$ and $m_y^2$ terms
in the modified eigenvalues $\tilde{\lambda}_m$ (Eq.\ (A13)),
and to a summation over all integers (including negative ones) for 
$m_x$ and $m_y$ in Eqs.\ (A15), (A16), (A24) and (A25).
The results are presented in Table \ref{table}.
The values are only half those given by Lee et al.\ due to a
factor $2$ in the definition of $g$. Also, our result for the
three--dimensional case is slightly higher, probably due to
our calculating the involved sums to a higher precision.
The boundary conditions have of course no effect for the quasi--one--dimensional
case. The values given here are those for the standard deviation
$\sigma_g$ for the orthogonal universality class of Random Matrix Theory.
The values for the other universality classes\cite{Pic91} are obtained by
dividing the variance, i.e.\ $\sigma_g^2$, by the universality
class parameter $\beta$, where $\beta = 1,2,4$ for the orthogonal, unitary,
and symplectic universality classes respectively.

\begin{figure}[h]
\resizebox{3.0in}{3.0in}{\includegraphics{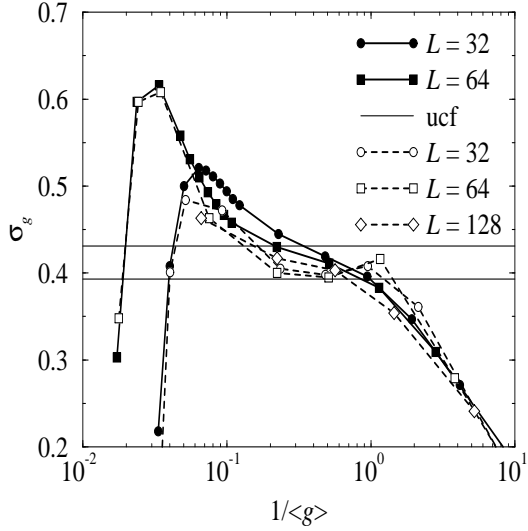}}
\caption{\label{2d-g}Standard deviation $\sigma_g$ of the conductance for
squares of $L\times L$ lattice sites. Full symbols: hard wall bc;
open symbols: periodic bc. Note that the bc seem to have little
effect outside the plateau region.The two horizontal lines indicate the
theoretical values for the UCF
for hard wall (top) and periodic (bottom) bc.}
\end{figure}

In Fig.\ \ref{2d-g} we show the standard deviation of the conductance
in ensembles of 10,000 samples for different system sizes (squares
with $L\times L$ lattice sites) and boundary conditions (open symbols: periodic boundary conditions;
filled symbols: hard wall boundary conditions) as a function of the inverse of the
average conductance. It is well known that in a two--dimensional disordered
tight--binding model all the states are exponentially localized.\cite{Eco84}
The typical structure\cite{Hig92,Tan94}
of the fluctuations with
increasing disorder strength can be seen: after an initial strong increase
in the ballistic regime (large $\left\langle g \right\rangle$) it
reaches a peak value which becomes more pronounced for larger
systems; then the fluctuations drop back to the universal value and
finally decrease again in the strongly localized regime.
The boundary conditions have apparently no influence on the behaviour 
outside the
region of UCF in this case.
Notice that for the case of periodic boundary conditions for the large
system size of $L = 128$, $\sigma_g$ approaches the theoretical value
of $0.393$ given by the lower horizontal line.

\begin{figure}[h]
\resizebox{3.0in}{3.0in}{\includegraphics{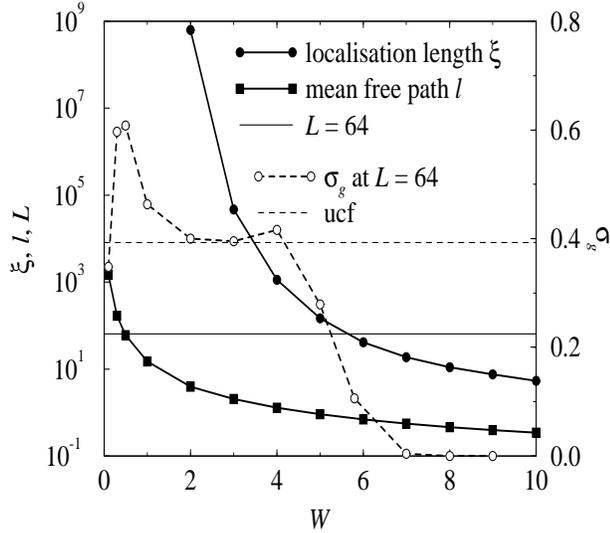}}
\caption{\label{2d-w}Standard deviation of a square of $64\times 64$
lattice sites together with the mean free path $l$ and the
localisation length $\xi$.}
\end{figure}

In Fig.\ \ref{2d-w} the same data is plotted for a two--dimensional
disordered system of size $L = 64$ with periodic boundary conditions.
In the same plot the mean free path and the localisation length 
as a function of disorder strength $W$ are given.
Both the mean free path and the localization length were obtained from
the numerical results of Economou et al.\cite{Eco84}
The localization length was obtained\cite{Eco84} by the transfer matrix method,
while the mean free path was obtained by the coherent potential
approximation\cite{Eco84} (CPA). Notice that $\xi$ is always larger than $l$.
So for a given system size ($L = 64$ in this case), there is a finite
region where $l \ll L \ll \xi$. Only in this region
there is a plateau visible at the correct UCF--value.
For $W \leq 1$, $l$ is larger than $L$, and we are in the ballistic regime
where one observes a monotonic increase of $\sigma_g$ followed by
the characteristic maximum as the system enters the crossover 
between the ballistic regime and the regime characterized by UCF.

\begin{figure}[h]
\resizebox{3.0in}{3.0in}{\includegraphics{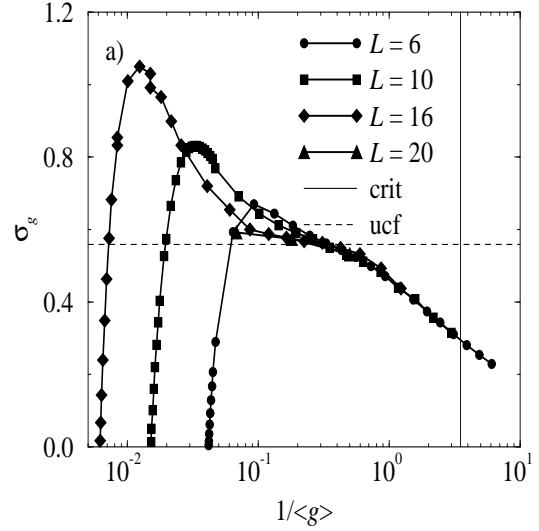}}
\resizebox{3.0in}{3.0in}{\includegraphics{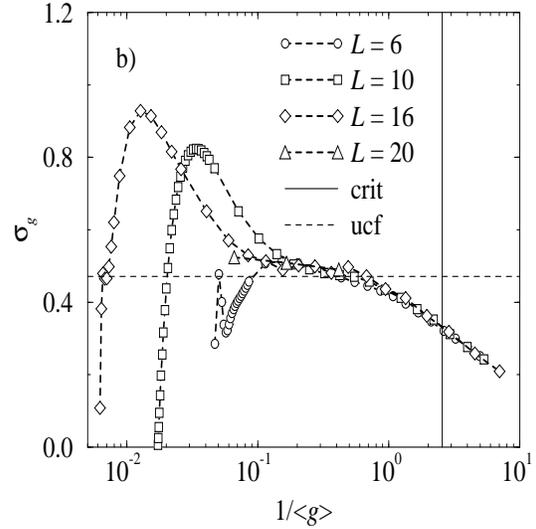}}
\caption{\label{3d-g}The standard deviation for systems of $L\times
L\times L$ lattice sites: a) hard wall bc; b) periodic bc.
The horizontal lines indicate the UCF values; the vertical lines
indicate the MIT.}
\end{figure}

Fig.\ \ref{3d-g} shows that the same overall behaviour is observed also
for three--dimensional systems. As there is a MIT (indicated by
vertical lines in Fig.\ \ref{3d-g}) where the conductance distribution
and therefore also its standard deviation become universal, i.e.\
independent of system size (though still depending on the boundary
conditions\cite{Sou99a,Sle00,Bra01}), this value is approached after leaving
the region of UCF. A direct comparison of the results shows again that
the boundary conditions have only minimal effect outside that region. The additional
peak noticeable in some of the periodic boundary conditions data are due to a
near--degeneracy of eigen--energies for very small disorder.

\begin{figure}[h]
\resizebox{3.0in}{3.0in}{\includegraphics{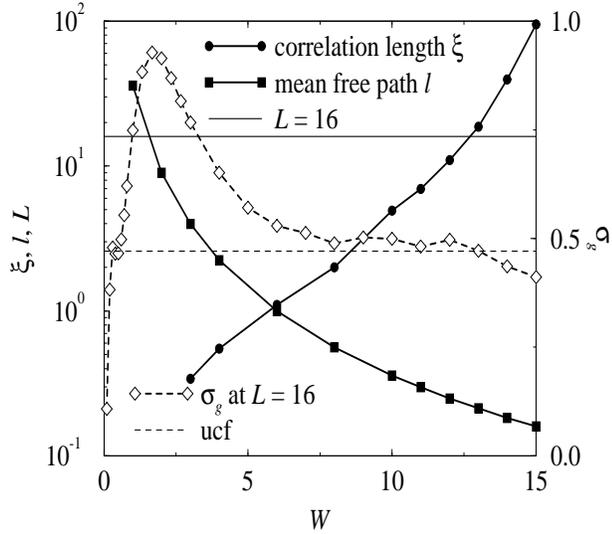}}
\caption{\label{3d-w}The standard deviation of a cube with $16 \times 16 
\times 16$ lattice sites together with the mean free path $l$ and the
correlation length $\xi$.}
\end{figure}

In Fig.\ \ref{3d-w} we plot the data for one of the systems again
as a function of $W$, together with the mean free path $l$
and the correlation length $\xi$. 
In the three--dimensional disordered case, the mean free path and the
correlation length $\xi$ were again obtained\cite{Eco85} by the CPA
and the transfer matrix method. The size of the cube is $L = 16$
and periodic boundary conditions were used. In the three--dimensional
case, $l$ drops as the disorder strength $W$ increases, while $\xi$
increases as $W$ increases. The plateau in $\sigma_g$ is seen only
when $l \ll L$ and $\xi \leq L$, so that we expect a {\it wider}
plateau for larger systems. The fluctuations begin to approach
the critical value as soon as $L \approx \xi$. In the three--dimensional
case too, when $W \leq 2$, $l$ is larger than $L$ and the ballistic regime
is observed, followed by the maximum in the crossover regime.

\begin{figure}[h]
\resizebox{3.0in}{3.0in}{\includegraphics{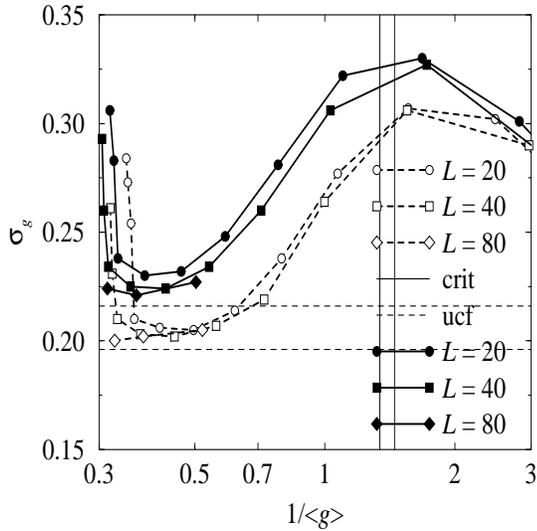}}
\caption{\label{2d-ev}The standard deviation for square systems of
$L\times L$ lattice sites with spin--orbit--interaction (Evangelou--Ziman
model) for hard wall (solid symbols) and periodic (open symbols) bc.
The two horizontal dashed lines indicate the theoretical values for the
UCF for hard wall (top) and periodic (bottom) bc.}
\end{figure}

Finally, Fig.\ \ref{2d-ev} shows data for square systems with 
spin--orbit--interactions. We have chosen the Evangelou--Ziman 
model,\cite{Eva87}
where even in the absence of diagonal disorder there is disorder
in the hopping matrices $V_{n,n'}$, which accounts for the fact
that the fluctuations do not vanish for small diagonal disorder.
Apparently, boundary conditions have a noticeable influence on the fluctuations
even outside the region of UCF, but this is likely due to the
peculiar overall structure in this case, most significantly the
fact that the UCF value is much smaller than the critical value,
causing another increase in the standard deviation as one
approaches the MIT.

In conclusion, we have investigated the conductance fluctuations
for various types of systems with both hard wall and periodic
boundary conditions from the ballistic regime to the localized
regime. The boundary conditions seem to have a relevant influence
on the conductance fluctuation only in the region of UCF and
at the critical point of the MIT. In true metallic systems, the
fluctuations begin to deviate from the UCF--value and approach
the critical value as soon as the correlation length approaches
the system size.

Ames Laboratory is operated for the U.S.\ Department of Energy by
Iowa State University under Contract No.\ W--7405--Eng--82.
This work was supported by the Director for Energy Research,
Office of Basic Science. The authors thank Professor E. N.
Economou for helpful discussion. P.M.\ would like to thank Ames Laboratory
for their hospitality and support and the Slovak
Grant Agency for financial support.

\begin{table}[h]
\caption{\label{table}The universal conductance fluctuation values
for different bc and dimensionality of the system.}
\begin{tabular}{|l||r|r|r|}
bc & Q1D & 2D & 3D\\ \hline\hline
hard wall & 0.365 & 0.431 & 0.559\\ \hline
periodic & 0.365 & 0.393 & 0.471\\
\end{tabular}
\end{table}

\end{document}